\newif\iffigs\figsfalse
\figstrue
\catcode`\@=11
\def\slash{\mathpalette\make@slash}
\def\make@slash#1#2{\setbox\z@\hbox{$#1#2$}%
\hbox to 0pt{\hss$#1/$\hss\kern-\wd0}\box0}
\catcode`\@=12 

\iffigs
\input epsf 

\else

\fi
\def\bbbz{{\bf Z}}
\def\bbbc{{\bf C}}
\def\bbbr{{\rm I\!R}} 

\def\bbb1{{\rm 1\!1}}

\newcommand{\eqnl}[2]{\par\parbox{11cm}
{\begin{eqnarray*}#1\end{eqnarray*}}\hfill
\parbox{1cm}{\begin{eqnarray}\label{#2}\end{eqnarray}}\break}

\newcommand{\eqngrlb}[3]{\par\parbox{11cm}
{\begin{eqnarray}\fbox{$\displaystyle#1\\#2$}\end{eqnarray}}\hfill
\parbox{1cm}{\begin{eqnarray}\label{#3}\end{\eqnarray}}\break}

\newcommand{\eqngrl}[3]{\par\parbox{11cm}
{\begin{eqnarray*}#1\\#2\end{eqnarray*}}\hfill
\parbox{1cm}{\begin{eqnarray}\label{#3}\end{eqnarray}}\break}

\newcommand{\refs}[1]{(\ref{#1})}
\def\PSL{P\!S\!L}

\def\apendix{\par
\setcounter{section}{0}
\setcounter{subsection}{0}
\def\thesection{Appendix \Alph{section}.}
\def\theequation{\Alph{section}.\arabic{equation}}
}

\documentstyle[12pt]{article}

\advance\hoffset by -.35in
\begin{document}
\bibliographystyle{unsrt}
\begin{flushright}
DIAS-STP/96-21 \\
BONN-TH-96-15\\
ENSLAPP-L-621/96\\
FAU-TP3-96/20\end{flushright}

\def\pr{\prime}
\def\pa{\partial}
\def\es{\!=\!}
\def\ha{{1\over 2}}
\def\>{\rangle}
\def\<{\langle}
\def\mtx#1{\quad\hbox{{#1}}\quad}
\def\pan{\par\noindent}
\def\La{\Lambda}

\def\A{{\cal A}}
\def\G{\Gamma}
\def\Ga{\Gamma}
\def\F{{\cal F}}
\def\J{{\cal J}}
\def\M{{\cal M}}
\def\R{{\cal R}}
\def\W{{\cal W}}
\def\tr{\hbox{tr}}
\def\al{\alpha}
\def\d{\hbox{d}}
\def\De{\Delta}
\def\L{{\cal L}}
\def\H{{\cal H}}
\def\Tr{\hbox{Tr}}
\def\I{\hbox{Im}}
\def\R{\hbox{Re}}
\def\ti{\int\d^2\theta}
\def\bti{\int\d^2\bar\theta}
\def\ttbi{\int\d^2\theta\d^2\bar\theta}


\begin{center}
\Large{{{\bf Uniqueness of the Seiberg-Witten Effective Lagrangian}}}
\end{center}

\begin{center}R. Flume\footnote{Volkswagen Foundation Fellow, on leave of 
absence University of Bonn}, M. Magro\footnote{On leave of absence from 
ENSLAPP, Groupe de Lyon, 46 all\'ee d'Italie, 69364 Lyon Cedex 07, 
France}, L.O'Raifeartaigh, I. Sachs and O. Schnetz\\
{\it Dublin Institute for Advanced Studies,\\
10 Burlington Road, Dublin 4, Ireland.}
\end{center}

\begin{abstract}
The low energy effective Lagrangian for $N\es 2$ supersymmetric Yang-Mills 
theory, proposed by 
Seiberg and Witten 
is shown to be the unique solution, assuming only that supersymmetry is 
unbroken and that the number of strong-coupling singularities is finite. Duality is 
then a consequence rather than an input.
\end{abstract}
\section{Introduction}
Over the last two years considerable progress in the 
understanding of the strong 
coupling regime in $N\es 2$ Yang-Mills theory has been made, 
pioneered by the work of Seiberg and Witten \cite{SW1} where 
a self-consistent non-perturbative 
superfield effective Lagrangian was found for an 
$SU(2)$ gauge group. This has 
later been extended to higher groups 
\cite{LAS}. The crucial 
properties which make the $N\es2$-theory accessible 
for an exact treatment are the holomorphic properties of the effective 
Lagrangian and the $1$-loop exactness of the 
perturbative $\beta$-function. The idea is 
then to extrapolate the perturbative (weak coupling) 
$\beta$-function to the full range of couplings. Making the 
assumption that the strong coupling behaviour is related in a definite way 
(ie. by $S$-duality) to 
the weak coupling regime lead to an elegant, self-consistent solution for 
the low energy effective Lagrangian in \cite{SW1}. Furthermore the duality is 
closely related to the electric-magnetic duality conjectured earlier by 
Olive and Montonen \cite{MO} to be a 
non-perturbative property of some quantum field theories. \par 
The authors of \cite{SW1} motivate the duality assumption with several 
convincing arguments based on physical intuition. Furthermore recent 
explicit $1$-and $2$-instanton computations 
\cite{Pouliot,Dorey,Ito,Fucito} 
confirm 
the first two coefficients in the asymptotic expansion of the exact 
solution 
proposed in \cite{SW1}. On the other hand strongly coupled systems 
have surprised us on several 
occasions with counter-intuitive results. Also, recent instanton calculations 
indicate that the extension of \cite{SW1} for models including 
matter multiplets given in \cite{SW2} is ambiguous \cite{Dorey1.5}.  
Therefore the question whether the assumptions made in \cite{SW1,SW2} are 
necessary or whether the exact effective Lagrangian can be obtained from 
weaker assumptions is of much interest. It is that question we address here. 
Indeed we prove that, provided supersymmetry is unbroken, the uniqueness of SW solution follows assuming only that the number of strong-coupling singularities on the moduli space is finite. 

Our strategy is as follows: we first construct the general effective 
Lagrangian, compatible with 
perturbation theory, analyticity and  the $\theta$-vacuum. 
Note that the first two conditions 
are consequences of supersymmetry alone. It turns out 
that the existence of the $\theta$-parameter restricts the set of 
admissible solutions considerably. Indeed the existence of the 
$\theta$-parameter already contains the seeds of duality, in the sense 
that either 
the theory has  complete 
$\PSL(2,{\bf Z})$-symmetry, or the moduli space  
contains a point which is conjugate to the weak coupling 
regime. 
Even so, the solution is not unique at that level. We then make use of the fact that for an asymptotically free theory, the scale of the low energy coupling is set by the mass of the lightest charged field. We therefore  impose the further constraint 
that the mass of the lightest charged field be finite except in the 
asymptotically free regime. The set of solutions then collapses to a 
single member, 
which is precisely the Seiberg-Witten solution. The same result is achieved by 
demanding that the expectation value $\Tr\<\phi^2\>$, where $\phi$ is 
the scalar component of the $N\es 2$-multiplet, is finite  except in the 
asymptotically free regime. This property  is very much expected for an 
asymptotically free theory. We also obtain the dependence of this 
observable on the point in the moduli space on general grounds. \par
The plan of the paper is as follows: In section $2$ we review the 
peculiar properties of $N\es2$ Yang-Mills theory which we will take as the 
only inputs for the later sections. In section $3$ we construct the general 
solution for the effective Lagrangian compatible with these requirements. 
In section $4$ we then include the finite mass constraint in our analysis and show that it reduces the above set of solutions to the SW one. The role of the expectation value $\Tr\<\phi^2\>$ is explained in section $5$. Some 
of the mathematical 
constructions needed for the main text and the argument relating 
the $\theta$-vacuum to duality are given in two appendices.  

\section{Review of $N\es2$ Yang-Mills}
A crucial property of $N\es2$-Yang-Mills theory is the presence of flat 
directions in the  potential for the scalar component $\phi$ of the 
$N\es2$-multiplet,  $V(\phi)\es\frac{1}{g^2}\mbox{Tr}[\phi,\phi^\dagger]^2$, 
where $g$ is the coupling constant. The potential vanishes for constant $\phi$ 
taking its value in the Cartan subalgebra of the gauge group. Furthermore, 
for unbroken supersymmetry,  
this degeneracy cannot be removed by quantum corrections \cite{W1}. For  $\phi\neq 0$, the Higgs 
mechanism breaks the gauge symmetry spontaneously down to 
$U(1)^l$,  where $l$ is the rank of the Cartan subalgebra. In what follows we concentrate on 
$SU(2)$. As explained in \cite{Sei,W3}, it can be deduced from $N\es 2$ supersymmetry 
that, when expressed in terms of the Cartan algebra-valued  $N\es2$-superfield 
$\A=\phi+\theta\chi+\cdots$, the most general local $N=2$ supersymmetric 
low energy effective 
Lagrangian 
must be of the form
\eqnl{\G[\A]=\frac{1}{4\pi}\hbox{Im}\int d^4x
d^2\theta_1 d^2\theta_2 \F(\A),}{eff1}
where the prepotential $\F$, to be determined, is the result 
of integrating out the massive (ie. charged, root-valued) fields. 
In $N\es1$ notation \refs{eff1} becomes \cite{G1} 
\eqnl{\G[A,W_\al]=\frac{1}{8\pi}\I\,\Tr\int\d^4x\Big\{ \ttbi 
(A_D\bar A-{\bar A}_D A)+\ti\tau(A)\;W^\al W_\al\Big\},}{action2}
where $A$ and $W_\al$ are the chiral- and vector $N\es1$ superfields 
respectively. Furthermore 
\eqnl{ A_D=\F'(A)\mtx{and} \tau(A)=\F''(A)=A_D'(A).}{ada} 
Since $\tau(A)$ is the coefficient of the kinetic term in \refs{action2} its 
imaginary part must be positive. On the other hand its real part plays the 
role of an effective 
$\theta$-angle: $\mbox{Re }\tau = \frac{\theta}{2 \pi}$. Thus a shift of 
$\theta$ by $2 \pi$ corresponds to 
$\tau \mapsto T(\tau) = \tau +1$. 
Therefore the group 
${\cal T}=\{T^n, n \in Z\}$ is a symmetry group of the theory. 
The invariance of the chiral part in \refs{action2} together with \refs{ada} 
requires then that 
${\cal T}$ be represented linearly on $(A,A_D)$ by a subgroup 
of $U(1) \times {\cal T}$. 

A remarkable consequence of $N=2$ supersymmetry is that the mass  
of the charged particles must be proportional to the central charge $Z$ of the SUSY-algebra. More precisely \cite{SW1,OW}
\eqnl{M=\sqrt{2}\vert Z\vert \quad \hbox{where} \quad Z={\bf n \cdot a}, 
\qquad {\bf a}=\pmatrix{a_D\cr a}\mtx{and} {\bf n}=(n_m,n_e)\neq0.}{Z}
Here $a=\Tr(\<\phi\> \sigma_3)$ is the expectation value of the scalar component of the $N\es2$ superfield and $a_D=\F'(a)$. The 
integers $n_e$ and $n_m$ label electric and magnetic charge 
respectively. The use of the dual variable $a_D$ exhibits the 
$S\!L(2,\bbbz)$ invariance of the mass formula \refs{Z} and of the 
first integral in \refs{action2}. \par
At the classical level $a_D=a \tau $. Furthermore the 
$U(1)$-invariance of the first integral in \refs{action2} reflects 
the $R$-symmetry of the Lagrangian. 
The theory is then parameterized by two real parameters, 
$g^2$ and $|a|$ (the phase of $a$ is irrelevant because of $R$-symmetry).\pan
In the quantum theory the mass of the charged fields
sets the scale for the low energy coupling. Because of asymptotic freedom
perturbation theory is then valid for large masses ($M\!>\!>\!\Lambda$). At the
semiclassical level we then have ($\Lambda\es 1$) 
\cite{Shif0} 
\eqnl{\tau(a)= \frac{i}{\pi} (\log (a^2) +c)}{tau}
where $c$ depends on the renormalization scheme adopted. 
The divergence of the $R$-current is 
in the same supermultiplet as the trace of the energy-momentum tensor and 
is therefore also anomalous at $1$-loop. Consequently $R$-symmetry is 
broken to the discrete group $a \rightarrow e^{\frac{i \pi}{2}} a$. Because 
of the $N\es2$ supersymmetry higher loop perturbative 
corrections to the running coupling are 
absent \cite{W2}. Note that due to the quantum corrections the low energy theory is 
parameterized by either of the complex parameters $a$ with values 
in $\bbbc$, or $\tau$ which takes any value in the upper half plane 
$H$. In particular, the space of inequivalent vacua, or moduli space, $\cal M$ is 
one (complex) dimensional.
\newpage
\section{Determination of $\F$}\label{strong}
\subsection{Statement of the Problem}
In addition to the perturbative corrections, reviewed in the last section, 
the low energy effective theory receives corrections due to topologically 
non-trivial configurations \cite{Afflek,Shif,Sei}. In principle, of course, 
$\F(a)$ could be computed directly from the functional 
integral but in practice this is far too difficult. The problem is then to 
determine $\F(a)$ or equivalently $\tau(a)$ from their properties established 
in the previous section: $\tau$ is an analytic function of $a$ with 
$\mbox{Im} (\tau(a)) \geq 0$ and satisfies the boundary condition 
\refs{tau}. The analyticity of $\tau$ follows from the analyticity of 
the prepotential $\F$. \par

The $S\!L(2,\bbbz)$ structure mentioned above and the structure of 
the instanton contributions to \refs{tau} found in \cite{SW1,Sei} 
suggest that, asymptotically at least, $\tau(a)$ is an 
inverse modular function. Seiberg and Witten proposed that $\tau$ be 
an inverse modular function everywhere on $\cal M$, where $\cal M$ is 
parameterized by some variable $u \in \bbbc$, with $u \rightarrow a^2$ 
asymptotically. To go further they made the two following 
assumptions:
\vskip 0.2truecm 
\noindent (i) {\it Minimality}:  $\tau(u)$ has just {\it two} 
singularities for finite $u$, 
\vskip 0.2truecm 
\noindent (ii) {\it Duality}: the monodromy matrix for 
$\tau(u)$ at one of the singularities is the transpose of that for 
$u\rightarrow \infty$.
\vskip 0.3truecm \noindent 
They  then showed that there is a unique function $\tau(u)$ with these properties, 
which in turn can be lifted to a unique effective prepotential $\F(a)$. 
Later it was 
found \cite{Pouliot,Dorey,Ito,Fucito} that the first two coefficients in 
the asymptotic expansion of the proposed $\F(a)$ agreed  with direct 
instanton computations. The latter result is evidently a strong 
indication that the SW Ansatz is correct. More recently it has been shown 
that $\F(a)$ can be obtained from somewhat weaker assumptions. However, a 
critical assumption 
that is made in \cite{Matone2} is that $\Tr(\<\phi^2\>)$ parameterizes 
the moduli space ${\cal M}$.  
As we shall see  
this is equivalent to assuming that the  
 Wronskian of $a$ and $a_D$ with respect to the moduli 
parameter is constant. \par
Below we show that these assumptions, although correct, are not necessary. 
Specifically our inputs are:
\vskip 0.2truecm 
\noindent (a) The effective coupling constant  
$\tau=\frac{\theta_{eff}}{2\pi}+i4\pi g_{eff}^{-2}$ takes all 
values in the upper half plane $H$.
\vskip 0.2truecm 
\noindent (b) The mass $m$ of the lightest charged field 
(possibly composite) is finite except in the asymptotically free region.
\vskip 0.2truecm 
\noindent (c) The mass $M$ is a single valued function on the moduli 
space,  $M=M(P)$, $P\in {\cal M}$.\pan
\vskip 0.2truecm 
\noindent (d) The set of singular points of ${\cal M}$ is finite.

\subsection{Uniformization}\label{uni}
Due to the $1$-loop corrections \refs{tau}, 
$Z$, $a_D$ and $\tau$ are multiple-valued transcendental functions of 
$a$, 
whereas $Z$, $a_D$ and $a$ are single-valued functions of 
the coupling 
$\tau \in H$. 
Thus,  
$\tau$ is the obvious candidate for the uniformizing parameter. Of course, 
distinct 
values of the effective coupling $\tau$ may correspond to equivalent vacua. 
For instance, we already know that $\tau$ and $\tau +1$ have to be 
identified. 
We therefore introduce the concept of a {\it maximal equivalence group} $G$.
This is defined as the group 
of all 
transformations $g$ of $\tau$ that leave $P(\tau)\in {\cal M}$ invariant\footnote{Strictly speaking, this group should be defined  
on the level of $\bf a$ rather than $\tau$; however, this distinction only becomes important 
when matter hypermultiplets  are included \cite{USW2}.} i.e. 
\eqnl{G=\{g \, / P(g\tau)=P(\tau), \; \forall \tau\in H\}.}{eg} 
The action of $G$ on $\tau$ then induces an action $\tilde G$ 
on ${\bf a}(\tau)$. 
The invariance of the spectrum \refs{Z} together with the integer valuedness 
of the charges $(n_m,n_e)$ imply that $\tilde G$ acts linearly on ${\bf a}$ 
and indeed  is a 
subgroup of $U(1)\times \PSL(2,\bbbz)$. 
The conditions $\I\tau \geq0$ and 
\eqnl{\tau(a)=\frac{\d a_D}{\d a},}{tau2}
then imply that $G$ is a subgroup of 
$\PSL(2,{\bf Z})\es S\!L(2,{\bf Z})/{\bf Z}_2$ and acts on $\tau$ by 
modular transformations. 
As noted in the context of the SW solution in \cite{Ferrari}, the $U(1)$ 
factor in $\tilde G$ cannot be ignored because 
the mass-spectrum is determined by $\vert Z \vert$ and not $Z$. As we shall 
see in the next section, and  was noted for the SW solution 
in \cite{Matone}, the $U(1)$-factor corresponds to the fact that 
${\bf a}$ is a section of a non-trivial $U(1)\times G$ bundle. \pan
We conclude that the set of 
inequivalent couplings $\tau$ is a fundamental domain 
$D=H/G$ of the maximal equivalence group $G$. The domain $D$ is then in 
$1\!-\!1$ correspondence with the moduli space ${\cal{M}}$. Because 
$G \subset \PSL(2,{\bf Z})$, $D$ is a polygon bounded by arcs 
\cite{Gunning}. \par
It is at this point that we make use of the assumption that the number 
of strong coupling singularities be finite. 
The corners of the fundamental domain $D$ correspond to singularities 
in the moduli space. Finite number of singularities then requires that the 
polygon $D$ has a finite number of corners. If this condition was not fulfilled then the set of corners of $D$ and therefore the set of singularities in ${\cal{M}}$ would necessarily have an accumulation point.

\subsection{General Solution}
In this section we show that for any group $G$ for which the fundamental domain has a finite number of corners, we can construct functions 
$a(\tau)$ and $a_D(\tau)$ such that $\I\tau \geq0$,
\eqnl{\tau = \frac{da_D}{da}\mtx{and}
\vert{\bf n \cdot a}(g\tau)\vert=\vert{\bf n' \cdot a}(\tau)\vert\mtx{for} 
g\in G.}{gta}
\paragraph{General Form of $\tau$:}
The polygon $D$ is bounded by arcs and has a finite number of corners. One of them corresponds to the weak-coupling singularity $\tau\es i\infty$.  
We now make use of the fact that $D$ can be parameterized  
by means of a Fuchsian mapping \cite{Magro} 
$\tau: z \in H\mapsto \tau(z) \in D$. The  Schwarzian of $\tau$ 
\eqnl{
\{\tau,z\}\equiv\frac{\tau'''}{\tau'}-\frac{3}{2}
(\frac{\tau''}{\tau'})^2}{schwa}
has the simple form
\eqnl{\{\tau,z\}=\sum\limits_{i=1}^n\big[\ha\frac{1-\al_i^2}
{(z-a_i)^2}+\frac{\beta_i}{z-a_i}\big],}{fuc1}
where $n+1$ is the number of edges of $D$, the 
$a_i$'s are the points on the real axis into which the 
corners of the polygon $D$ are mapped. Furthermore the $\alpha_i\in[0,1)$ 
are the 
interior angles 
of the polygon. In \refs{fuc1} we have chosen to map 
the weak-coupling singularity 
$\tau = i\infty$ to infinity in the 
$z$-plane. Since the corresponding angle is zero we have \cite{Magro}
\eqnl{\{\tau,z\} \rightarrow \frac{1}{2z^2}\mtx{for} 
z\rightarrow \infty.}{compact} 
This condition puts two constraints on the 
$\beta$'s. 
Thus there are $3n-2$ parameters for each set of 
polygons with $n$ corners. Actually only $3n-4$ of these are 
independent because the origin and the scale  of $z$ on the real axis are 
still free parameters.  If we furthermore identify segments on the real line in the 
$z$-plane, that are mapped into edges 
of the polygon $D$ which are equivalent with respect to $G$, then the 
equivalence group $G$ of $\tau$ 
is just its monodromy group with respect 
to $z$. This makes it natural to  parameterize the moduli 
space by $z\in H$. \pan 
The useful property of Fuchsian functions is that $\tau(z)$ can be 
written as \cite{Magro}
\eqnl{\tau(z)=\frac{y_1(z)}{y_2(z)},}{ty}
where ${\bf y} = (y_1,y_2)$ is solution of the second order 
differential equation ($'=\frac{d}{dz}$)
\eqnl{{\bf y}''+Q{\bf y}=0\mtx{with} Q(z)=\ha\{\tau,z\}.}{ydiff}
Since the Wronskian $W(y_1,y_2)=y'_1y_2-y_2'y_1$ of $y_1$ and $y_2$ 
is constant $G$ acts linearly on ${\bf y}$ with a trivial $U(1)$. 

\paragraph{Lifting:}
We now show that the previous second order differential equation 
for ${\bf y}$ can 
be lifted to a second order differential equation for ${\bf a}$. 
This may come as a surprise as the lifting from $\tau$ to $(a_D,a)$ involves 
an integration leading {\it a priori} to a third order differential equation 
for ${\bf a}$. 
However, we show below that it is always possible to lift \refs{ydiff} 
to a second order differential equation for ${\bf a}$ provided 
we allow for an extra $U(1)$-factor to appear in the action of $G$ on  
${\bf a}$. \par
Define ${\bf a}$ by
\eqnl{{{\bf a}}=f'{{\bf y}}-f{{\bf y}}'={{\bf W}}(f,{{\bf y}}),}{fy}
where $f$ is any section of the $U(1)$-bundle and ${\bf W}(f,{{\bf y}})$ 
is the 
Wronskian of $f$ and ${\bf y}$. Differentiating \refs{fy} with 
respect to $z$ gives 
\eqnl{{{\bf a}}'=f''{{\bf y}}-f{{\bf y}}''=(f''+Qf){{\bf y}}}{va}
and thus ${\bf a}$ satisfies \refs{gta}. Furthermore it is easy to see 
from \refs{ydiff} 
and \refs{fy} that ${\bf a}$ satisfies a second order 
differential equation.\pan
The construction \refs{fy} is in fact the only way to satisfy \refs{gta}. 
Indeed, from \refs{gta} we conclude that
\eqnl{{\bf a'}=g{\bf y},}{atoy}
where $g$ is some $U(1)$-section. 
The integral of \refs{atoy} is precisely given by  \refs{fy} with $f$ 
satisfying 
the inhomogeneous form of \refs{ydiff}
\eqnl{f''+Qf=g.}{fg}
The  constant of 
integration 
must be set to zero for ${{\bf a}}$ to transform homogeneously with respect to 
$G$. Note that the action of $G$ on ${\bf a}$ is also that of monodromy 
transformations in $z$. \par
\paragraph{Boundary Conditions:}
To complete the construction we need to satisfy the boundary conditions 
given by the semiclassical contribution \refs{tau}. We recall from 
\refs{compact} and \refs{ydiff} that
\eqnl{Q(z)\simeq\frac{1}{4z^2}\mtx{for} z\rightarrow\infty.}{assQ}
Therefore
\eqnl{{\bf y}\propto z^{{1 \over 2}}(1, \hbox{ln}(z)) 
\mtx{for} z\rightarrow \infty}{assy2}
and hence  
\eqnl{\tau(z)\rightarrow {ic\over \pi}\hbox{ln}(z).}{asstau}
The constant $c$ is fixed by using the fact that the action of $G$ 
on $\tau$ corresponds to monodromy transformations in $z$ 
and by requiring that 
$T:\tau\mapsto\tau+1$ belongs to $G$ as shown in section $2$.  First 
due to the identifications on the real line, for 
large $z$, $z \mapsto e^{i\pi}z$ is  a monodromy transformation. Under such 
a transformation $\tau \mapsto \tau +c$. Therefore $c$ is an integer. The 
condition $T\in G$ leads then to $c=1$.  
Compatibility with the semiclassical relation \refs{tau} then 
requires $a\propto z^\ha$ for large $z$, which using \refs{fy} and 
\refs{assy2} leads to
\eqnl{f(z)\propto z\mtx{for} z\rightarrow \infty.}{fg2}
This completes the general construction of the vector ${\bf a}(z)$. 
To summarize we have 
shown  that the action of any equivalence group $G$ for $\tau$ can be lifted 
to the pair of functions $a_D(\tau)$ and $a(\tau)$ and furthermore the action 
is simply by monodromy transformations on $z\in H$ with proper 
identifications. Without further constraints the solution is in general not 
unique. Indeed any Fuchsian function $\tau(z)$ mapping $H$ into a 
fundamental domain of $G$ and any $U(1)$-section $f(z)$ satisfying the 
boundary condition \refs{fg2} is a solution.

\section{Finite-Mass Constraint}\label{fmass}
As discussed in section $2$, infinite mass for all 
(possibly composite) charged fields implies that the full $SU(2)$-theory 
is weakly coupled due to asymptotic freedom. Therefore consistency requires 
that the mass $m$ of the lightest charged field diverges only 
in the perturbative regime. It turns out that the finite mass constraint puts 
a very strong condition on the set of solutions constructed in the 
last subsection. It is here also that the extra 
$U(1)$-bundle introduced by the function $f$ in 
\refs{fg} becomes important. 

\subsection{Finite-Mass Condition for $f$}
We first observe that in order to be a $U(1)$-section $f(z)$ must behave 
in the vicinity of a singularity $z_{0}$ of $f$ as 
\eqnl{f\left( z-z_{0}\right) 
\propto\left( z-z_{0}\right)^{r_{z_{0}}},}{floc}
(at infinity $ f(z)=1/z^{r_{\infty }}$). 

We now derive a lower bound for each $r_{z}$ from the finite mass condition. 
First we recall that the mass spectrum is given by 
$M=\sqrt{2}\vert{\bf n \cdot a}\vert$ and thus using \refs{fy} 
$M=\sqrt{2}\vert{\bf n \cdot W}(f,{\bf y})\vert$.
Finiteness of the mass spectrum 
requires then that at least one component of ${\bf W}(f,{\bf y})$ 
be finite at a given finite point $z_0$. \pan 
Now, if $Q$ is regular at this point  
then $y_1$ and $y_2$ are also regular and then finiteness of 
$M$ implies either $f$ is regular at $z_0$ or $f$ has a singularity at $z_0$ 
with $r_{z_0}\geq 1$. If $z_0$ is a singularity of $Q$, 
ie. $z_0=a_i$,
then \refs{ydiff} can be solved locally to give for $\alpha_i\neq 0$
\eqnl{y_\pm(z) \propto (z-a_i)^{\ha(1\pm\alpha_i)},}{diag} where $y_\pm$ are 
the eigenvectors of the monodromy matrix $M_{a_i}$.  Since
\eqnl{W(f,y_\pm)(z)\simeq c \Big(r_{a_i}-\ha(1\pm\al_i)\Big)
(z-a_i)^{r_{a_i}-\ha(1\mp\al_i)},\,\,\, c=\mbox{const}}{W3}
finiteness of the mass implies 
\begin{equation}
r_{a_i}\geq\ha(1+\al_i)\,\,\, \mbox{or} \,\,\, r_{a_i}=\ha(1-\al_i).\label{coa}
\end{equation}
For $\alpha_i=0$, $y_1$ and $y_2$ are linear 
combinations of $z^{\frac{1}{2}}$ and $z^{\frac{1}{2}} \log z$ and the 
same analysis gives $r_{a_i} \geq \ha$. Thus the condition \refs{coa} is 
also valid for this case. Combining the above results we see that {\it necessary} conditions for finite $M$ are
\eqngrl{r_{z} \geq\ha(1-\al_i)&\mtx{for}& z=a_i}
{r_{z}\geq1&\mtx{for}& z=b_i,}{nes}
where $\{a_i\}$ are the common singularities of $f$ and $Q$ and  
$b_i\not\in\{a_i\}$ are the points where $f$ is singular but 
$Q$ is regular.
\subsection{Total Residue Condition}
To proceed further we need a general theorem on residues, namely 
that the total residue i.e. the sum of the residues \cite{cohn} 
of a meromorphic form $\omega$ on a compact manifold is zero,
\begin{equation}
\sum_i \oint_{C_i} \omega = 0,
\end{equation}
where the integrations 
are taken along closed curves $C_i$ associated with any triangulation of the manifold.
Applying this result to the moduli space $\cal M$, the form 
$\frac{df}{f}$, which is meromorphic because $f$ is a section of a line bundle, and a triangulation of $\cal M$ around all inequivalent poles 
and zeros of $f$, we get
\begin{equation}
\sum\limits_{z{\rm \ interior\hspace{.38ex}}}r_{z}+\frac{1}{2}
\sum\limits_{z\in\bbbr}r_{z}+\ha r_{\infty }=0.\label{cc}
\end{equation}
The $\frac{1}{2}$'s in \refs{cc} are 
due to the fact that singular points on the real line are 
pairwise identified.


\subsection{Resolution of the Finite Mass Condition}
An immediate consequence of \refs{cc} is that $f$ cannot have a singularity 
at a point where $Q$ is
regular because a single singularity of this kind would already saturate
\refs{cc} with 
$r_\infty\es -1$ from \refs{fg2} and $\tau(z)$ has 
at least two singularities. Hence we can restrict 
ourself to the first 
possibility in \refs{nes}. Using \refs{cc} we then obtain
\eqnl{n-\sum_{i=1}^{n}\alpha_i\leq 2.}{rsum}
The l.h.s. of  
\refs{rsum} is directly related to the index $\mu$ of $G$ in 
$\PSL(2,\bbbz)$. The index $\mu$ is the order of the 
coset $\PSL(2,{\bf Z})/G$. To see the connection 
we endow the upper half plane $H$ with 
the $ \PSL(2,{\bf Z})$-invariant 
metric \cite{Frakas}
$\left( {\rm\hspace{.38ex}Im\hspace{.38ex}}\tau\right) ^{-2}{} 
d\tau d\bar{\tau}\hspace{.6ex}.$ 
With this metric every copy of the fundamental domain $D_0$ 
of $\PSL(2,{\bf Z})$ 
has the same area $ \pi /3$. Since the fundamental domain $D$ is composed 
of $\mu $ copies of $D_0$ \cite{Gunning} it has the area $ \mu \pi /3$. On the other 
hand the area of  $D$, which is a polygon bounded by arcs with centers 
on the real line 
and which has one zero angle at infinity and 
$n$ further angles $\alpha _{i}$ is \cite{Frakas} 
$\pi (n-1 -\sum_{1}^{n}\alpha_i)$.
The equality of the two expressions obtained for the area of $D$ leads to 
the relation   
\begin{equation}
n-\sum _{i=1}^{n}\alpha _{i} =\frac{\mu}{3}+1. 
\end{equation}
Thus the condition \refs{rsum} is equivalent to $\mu \leq 3$. 
It is shown in Appendix $A$, that 
the only subgroups of $ \PSL(2,{\bf Z})$ with index not greater than 
3 and containing $T$ are $ \Gamma _{0}(2)$ and $ \PSL(2,{\bf Z})$ itself. 
$\PSL(2,{\bf Z})$ is ruled out for the following reason.
Since $\alpha _{1}=\alpha _{2}=1/3$ the only $f$ consistent with \refs{cc} 
has $ r_{a_{1}}=1/3$, $ r_{a_{2}}=2/3$, or vice versa. However, 
since $a$$_{1}$ and $a$$_{2}$ correspond to couplings $\tau$ that are 
identified by $T$, $f$ cannot have different indices at $a$$_{1}$ and 
$a$$_{2}$.\pan
For $\Gamma_0(2)$ there are two sets of 
$\alpha_i$ and $r_z$:
\begin{eqnarray}
\{\alpha_1=\alpha_2=0\}&f(z)=c(z^2-1)^\ha, \label{FFF}\\
\{\alpha_1=\alpha_2=\ha,\alpha_3=0\}&
f(z)=c(z^2-1)^{\frac{1}{4}}(z-a_3)^{\ha},\label{fff}
\end{eqnarray}
where we have chosen two singularities $a_1$ and $a_2$ to be $1$ and $-1$.
These two  solutions correspond to two different choices of the domain $D$ 
and are related by a coordinate transformation  
on the moduli space. It can be checked that the physical quantities such 
as the mass spectrum are scalars under this transformation.\pan
The $\Gamma_0(2)$ solution just discussed is the SW solution. 
This may be surprising 
since the group found in \cite{SW1} is $\Gamma(2)$, whose index in 
$\PSL(2,\bbbz)$ is $6$. 
This difference comes only from the fact that in SW description $T$ plays 
a distinct role and is not associated to a monodromy transformation of their 
moduli parameter $u$. In other words $\Gamma(2)$ is not the {\it 
maximal} equivalence group but a subgroup of it. More precisely,  
$\Gamma(2)=\Gamma_0(2)/\bbbz_2$. 
The link between the $\Gamma_0(2)$ and SW descriptions  
will be given  more explicitly at the end of the next section. 

\section{Physical Description of $\cal M$}\label{uvar}

In this section we wish to consider the expectation value
\begin{equation}
\hbox{Tr}(\<\phi^2\>) \label{tphi}
\end{equation}
which, like $M$ has a direct physical meaning. In \cite{Fucito,Dorey2} it was 
shown that within the instanton approximation 
\eqnl{\hbox{Tr}(\<\phi^2\>)=\pi i (\F(a)-\ha a\F'(a)).}{tF}
Later it was shown in \cite{Howe} that \refs{tF} was generally true 
as a direct consequence 
of the superconformal Ward identities.  
In \cite{SW1} Seiberg and Witten conjectured that their 
moduli variable $u$ was identical with $\hbox{Tr}(\<\phi^2\>)$. 
That this is correct can be seen using the result 
\cite{Matone} which precisely states that the right hand side of \refs{tF} 
equals $u$. For this reason and for brevity we define  
$u\equiv\hbox{Tr}(\<\phi^2\>)$.
We now discuss the relation between $u$ and the moduli parameter $z$ in the 
general setting of the previous sections.  For this we 
note from \refs{tF} that 
\begin{equation}
1=\frac{i \pi}{2}(a_D \frac{\partial a}{\partial u} -a \frac{\partial a_D}
{\partial u}) \quad \Rightarrow \quad  \frac{du}{dz} = \frac{i \pi}{2}
(a_Da'-aa_D').
\end{equation}
Using \refs{atoy} we then have 
\eqnl{\frac{du}{dz}=\frac{i\pi}{2} f^2(\frac{f''}{f}+Q)(y_1y_2'-y_2y_1')=
f^2(\frac{f''}{f}+Q),}{I2}
where we have used \refs{fy} and normalized the Wronskian of $y_2$ 
and $y_1$ to be $\frac{2}{i\pi}$. 
In order to obtain $u(z)$ we need to integrate this identity  ie.
\eqnl{u(z)-u(z_1)=\int\limits_{z_1}^z f^2(z')(\frac{f''}{f}
+Q)(z')dz'.}{I3}
This is the required relation between $u$ and $z$. \par
Next we show that the Seiberg-Witten solution is the unique solution for 
which $u$ is finite except for the asymptotic regime, even if we 
do not insist on the finite mass constraint. For this we note that, 
 using \refs{fuc1}, the expression \refs{I3} is locally of the form 
\begin{equation}
\left\{\begin{array}{ccc}
\int\limits^z (z'-a_i)^{2r_{a_i}-2}
\Big[ \frac{1}{4}(1-\al_i^2)
+r_{a_i}(r_{a_i}-1)\Big]dz'&\mbox{for}& z \rightarrow a_i\\
\\ \label{ft}
\int\limits^z r_{z_0}(r_{z_0}-1)(z'-z_0)^{2r_{z_0}-2}dz'&\mbox{for}&
z\rightarrow z_0 \notin\{a_i\}
\end{array}\right\}
\end{equation}
where we have used $f(z)\propto (z-z_0)^{r_{z_0}}$. 

Finiteness of $u$ then leads 
to the following conditions: 
\vskip 0.2truecm 
\noindent (i) To every singularity of $Q$ corresponds a zero of $f$. 
Furthermore, either $r_{a_i}>\ha$ or $r_{a_i}=\ha(1-\al_i)$. 
\vskip 0.2truecm 
\noindent (ii) If $f$ has a singularity where $Q$ is regular ie. 
$z_0\not\in\{a_i\}$, then $r_{z_0}>\ha$. \pan
These conditions are rather similar to those obtained from the 
finite mass constraint. which is, perhaps, not too surprising because of 
the identity 
\eqnl{{dZ \over du}={{\bf n \cdot y} \over f} }{8.4}
which follows immediately from \refs{I3} and \refs{atoy} \refs{fg}.
However, they differ in two small respects. First 
(i) permits a small range $\ha(1+\alpha_i)> r_{a_i}>\ha$ that is not 
permitted by 
the mass condition. Second (ii) requires $r_{z_0}>\ha$ rather than 
$r_{z_0}\geq1$. However, 
when 
we apply the boundary condition \refs{fg2} we find that the only solution 
permitted by (i) and (ii) which is not permitted by the finite mass 
constraint is 
\eqnl{r_{a_1} = r_{a_2} = \frac{1}{6},\,r_{z_0}=\frac{2}{3}\mtx{and}
\al_{1}=\al_2=\frac{2}{3}.}{ce}
It is easy to see, however, that the map \refs{ce} cannot correspond to any 
subgroup of $\PSL(2,{\bf Z})$. 
Thus $u<\infty$ leads to exactly the same result as the finite mass 
constraint and therefore leads to a unique solution which is precisely the 
SW-solution. It then follows that $u\es\hbox{Tr}(\<\phi^2\>)$ is a good parameter for the moduli 
space. However, in this paper, in contrast to \cite{SW1,Matone}, this 
property of $u$ is not an 
assumption but a 
consequence of the finiteness of $m$ or $u$ itself. \par

Finally, for $\Gamma_0(2)$ the relation \refs{I2} can be explicitly 
integrated for the two cases \refs{FFF} and \refs{fff}. 
It leads respectively to 
$u=z$ and $u=\sqrt{1-z^2}$. Moreover it follows then from \refs{ydiff} 
and \refs{fy} that for both cases ${\bf a}(u)$ satisfies the second 
order differential equation 
\eqnl{\frac{d^2{\bf a}}{du^2} + \frac{1}{4(u^2-1)}{\bf a}=0},
which is just the differential equation obtained in \cite{Bilal} 
for SW solution.

\section{Conclusions}

We have proved that Seiberg-Witten Ansatz is the unique solution for the 
low energy effective Lagrangian of $N\es2$-Yang-Mills theory, assuming only 
that supersymmetry is unbroken and that the number of singularities is finite.
In particular the electromagnetic duality is derived,  
in contrast to the original paper \cite{SW1} where it was 
assumed. \pan
{\it En route,} 
we have obtained a construction which lifts any  
$\PSL(2,\bbbz)$-structure. This construction is straightforward and based on 
simple differential equations. In particular it does not involve elliptic 
curves. On the other hand it generalizes the observations in 
\cite{Bilal,Matone} and more recently \cite{Nahm} where the connection 
with differential 
equations was explained for the particular case of the Seiberg-Witten 
solution. \pan  
There is no reason why the present construction should not apply to theories 
including matter hypermultiplets. It therefore has the potential to 
resolve the puzzles arising there 
\cite{Dorey1.5}. The question of whether it can be extended to higher 
groups, where the complex dimension of the moduli space is bigger than $1$ 
is, however, still open.   

\section*{Acknowledgments}
I.S. was partially supported by the Swiss National Science 
Foundation. M.M. is grateful to the Dublin Institute for Advanced 
Studies for their kind hospitality. O.S. was supported 
in parts by the DFG Graduiertenkolleg `Starke Wechselwirkung' 
and the BMBF.  R.F acknowledges the support of
Volkswagenstiftung. 
I.S. is indebted to F. Hirzenbruch for helpful comments on modular groups.

\apendix

\setcounter{equation}{0}

\section{Modular Groups of Index $\mu\leq3$}
In this Appendix we prove some results on modular groups 
needed in the main body of the 
paper. First we recall some basic definitions and results \cite{Gunning}.\pan
A modular transformation, is a transformation
\eqnl{z\rightarrow\frac{az+b}{cz+d}\mtx{where}\pmatrix{a&b\cr c&d\cr}\in 
S\!L(2,\bbbz).}{fra}
It is called {\it elliptic, hyperbolic} or {\it parabolic} corresponding 
to 
whether it has two fixed points in the upper half plane, two fixed points 
on 
the real axis or one fixed point at $\infty$ or on the real line. These 
three 
cases can equivalently be distinguished by the trace (smaller, bigger or 
equal to $2$).\pan
For our purpose we restrict ourself to the inhomogeneous modular group
\eqnl{\PSL(2,{\bf Z})=S\!L(2,{\bf Z})/\pm I \quad \mbox{generated by} \quad 
\{T,S\} \quad \mbox{with} \quad \Big\{\begin{array}{l}
T: \tau \mapsto \tau +1 \\
S: \tau \mapsto -\frac{1}{\tau}
\end{array}.}{mod}
$H/\PSL(2,{\bf Z})$ is isomorphic to the Riemann sphere. More generally,
if $G$ is a subgroup of $\PSL(2,{\bf Z})$ of finite index, 
its fundamental domain 
${\overline {H/G}}$ is isomorphic to a compact Riemann surface \cite{Gunning}.\pan
Next we prove the following theorem used in sections \ref{fmass} and 
\ref{uvar}. \pan 
{\bf Theorem} \ {\it The only subgroups of 
$ \PSL(2,{\bf Z})$ with index less or equal 3 and containing 
$T$ are:
\begin{eqnarray*}
PSL\left( 2,{\bf Z}\right) &\quad& \mbox{generated by} \quad \{ T,S\} \\ 
\Gamma _{0}\left( 2\right) &\quad& \mbox{generated by} \quad \{T,ST^{2}S\} 
\quad \mbox{with} \quad ST^2S:\tau \mapsto 1/
\left( -2\tau +1\right)
\end{eqnarray*}}

\vspace{1ex}
\noindent{}

\noindent {\bf Proof:} First we recall \cite{Gunning} that $S^{2}=(ST)^{3}=1$.
If $S\in G$ then $G=\PSL(2,{\bf Z})$ otherwise 
$GS$ defines one coset of $G$. In this case $ST\notin G$. Moreover 
$ST\notin GS$, otherwise $G\ni STS=T^{-1}ST^{-1}$, thus $S 
\in G$. So $ GST$ defines another coset of $G$ and $G$ is of index $3$. 
Consider $STS$. It has to belong to one of the cosets $G$, $GS$ or $GST$.
But $STS\notin G$, otherwise $ TSTST=S\in G$. Moreover $ STS\notin GS$. Hence 
$STS\in GST \Rightarrow G\ni STST^{-1}S=ST^{2}ST$. Therefore $ST^{2}S\in G$ 
and $G=\Gamma _{0}(2)$.
\hfill $\Box$
\iffigs

\begin{figure}[h]\caption{Fundamental domains for subgroups of the 
modular group}
\input epsf
\epsfxsize=7cm
\epsffile{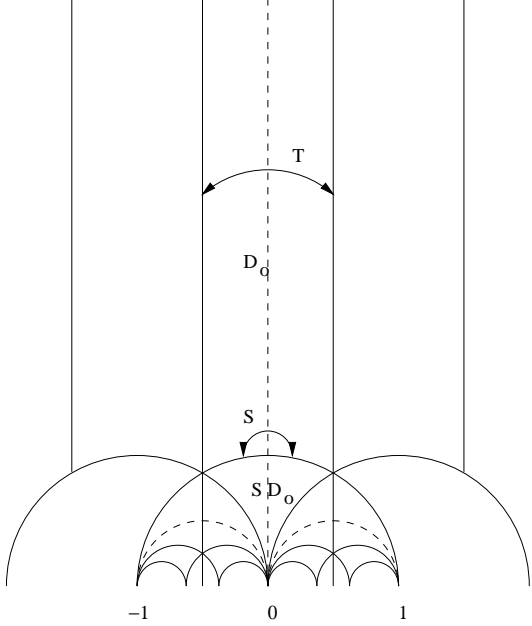}
\end{figure}
\vspace{1cm}

\else
\message{No figures will be included. See TeX file for more
information.}
\fi
\section{$\theta$-Vacuum and Duality}
In order to discuss the relation between the $\theta$-vacuum and duality, we 
need the following result on fundamental domains:\pan
{\bf Lemma} {\it Let $G\subset \PSL(2,{\bf Z})$, $G\neq \PSL(2,{\bf Z})$ 
such that 
$T\in G$. 
Then every fundamental domain $D$ of $G$ has at least one vertex on 
the real line.}\pan
{\bf Proof:}
Let $D_0$ be the usual fundamental domain of $\PSL(2,\bbbz)$ given 
in Figure 1. 
Because $G\neq \PSL(2,{\bf Z})$, it follows that $S\not\in G$. Therefore $D$ 
contains $SD_0$ or a copy of $SD_0$ obtained by 
applying some element $g$ of $G$ to $SD_0$. The image of $0\in SD_0$ 
under a modular transformation is either on 
the real line or at $i\infty$. However, the latter possibility leads to 
$gSD_0=T^nD_0$ for some integer $n$ which implies again that 
$G=\PSL(2,\bbbz)$. Hence we conclude that $D$ has 
a 
vertex on the real line. \hfill $\Box $\pan

Vertices on the real line always correspond 
to zero angles ie. parabolic substitutions ie. trace $2$.
>From the above it then follows in particular that any solution for the 
low energy effective Lagrangian which respects the existence of the 
$\theta$-parameter, either has 
the monodromy group $\PSL(2,{\bf Z})$ or otherwise the fundamental domain has 
at least $1$ parabolic vertex in addition to the parabolic point at infinity. 
The corresponding monodromy, does not commute with the 
monodromy at infinity. On the other hand it is conjugate 
(in $\PSL(2,\bbbz)$) to the point at infinity. Consequently, there is at 
least $1$ point in the vacuum manifold which is dual (conjugate)  to 
to the weak coupling singularity.\par

\end{document}

\end